\documentclass[final,3p,times]{elsarticle}
\pdfoutput=1

\usepackage[english]{babel}
\usepackage{graphicx}
\usepackage{amssymb}
\usepackage{subfig}
\usepackage{soul}
\usepackage[utf8x]{inputenc}

\usepackage{lineno}

\biboptions{sort&compress}

\newcommand{\nima}        [3]  {{\sl Nucl. Instr. and Meth. in Phys. Res. A}~{\bf #1} (#2) #3}
\newcommand{\itns}        [3]  {{\sl IEEE Trans. Nucl. Sci.}~{\bf #1} (#2) #3}
\newcommand{\jinst}       [3]  {{\sl Journal of Instr.}~{\bf #1} (#2) #3}
\newcommand{\rmp}         [3]  {{\sl Rev. Mod. Phys.}~{\bf #1} (#2) #3}

\newcommand{\epjc}        [3]  {{\sl Eur. Phys. J. C}~{\bf #1} (#2) #3}

\newcommand{\etal} {\emph{et.~al.}}

\journal{Nuclear Instruments and Methods in Physics Research: Section A}

\begin{document}

\begin{frontmatter}

\title{A compact light readout system for longitudinally segmented shashlik calorimeters}

\author[univ,infn_mib]{A.~Berra\corref{corr}}
\author[univ,infn_mib]{C.~Brizzolari}
\author[infn_bo]{S.~Cecchini}
\author[infn_bo]{F.~Cindolo}
\author[iphc]{C.~Jollet}
\author[infn_lnf]{A.~Longhin}
\author[infn_rm]{L.~Ludovici}
\author[infn_bo]{G.~Mandrioli}
\author[infn_bo]{N.~Mauri}
\author[iphc]{A.~Meregaglia}
\author[infn_lnf]{A.~Paoloni}
\author[infn_bo,univ_bo]{L.~Pasqualini}
\author[infn_bo]{L.~Patrizii}
\author[infn_bo]{M.~Pozzato}
\author[infn_lnf]{F.~Pupilli}
\author[univ,infn_mib]{M.~Prest}
\author[infn_bo]{G.~Sirri}
\author[infn_mib,univ_mib]{F.~Terranova}
\author[infn_ts]{E.~Vallazza}
\author[infn_lnf]{L.~Votano}

\address[univ]{Universit\`a degli Studi dell'Insubria, Via Valleggio, 11 - 22100 Como, Italy}
\address[infn_mib]{INFN sezione di Milano Bicocca, Piazza della Scienza, 3 - 20126 Milano, Italy}
\address[infn_bo]{INFN sezione di Bologna, Viale Berti Pichat, 6/2 - 40127 Bologna, Italy}
\address[iphc]{IPHC, Universit\'e de Strasbourg, CNRS/IN2P3, Strasbourg, France}
\address[infn_lnf]{INFN Laboratori Nazionali di Frascati, Via Enrico Fermi, 40 - 00044 Frascati (RM), Italy}
\address[infn_rm]{INFN sezione di Roma, Piazzale Aldo Moro, 2 - 00185 Roma, Italy}
\address[univ_bo]{Universit\`a di Bologna, Dipartimento di Fisica, Via Irnerio, 46 - 40126 Bologna, Italy}
\address[univ_mib]{Universit\`a degli Studi di Milano Bicocca, Dipartimento di Fisica, Piazza della Scienza, 3 - 20126 Milano, Italy}
\address[infn_ts]{INFN sezione di Trieste, Via Valerio, 2 - 34127 Trieste, Italy}

\cortext[corr]{\scriptsize{Corresponding author, address: Universit\`a degli Studi dell'Insubria, Via Valleggio, 11 - 22100 Como, Italy; e-mail: alessandro.berra@gmail.com}}

\begin{abstract}

The longitudinal segmentation of shashlik calorimeters is challenged by dead zones and non-uniformities introduced by the light collection and readout system. This limitation can be overcome by direct fiber-photosensor coupling, avoiding routing and bundling of the wavelength shifter fibers and embedding ultra-compact photosensors (SiPMs) in the bulk of the calorimeter. We present the first experimental test of this readout scheme performed at the CERN PS-T9 beamline in 2015 with negative particles in the 1-5~GeV energy range. In this paper, we demonstrate that the scheme does not compromise the energy resolution and linearity compared with standard light collection and readout systems. In addition, we study the performance of the calorimeter for partially contained charged hadrons to assess the $e/\pi$ separation capability and the response of the photosensors to direct ionization.
\end{abstract}

\begin{keyword}
Shashlik calorimeter \sep Longitudinal segmentation \sep Silicon PhotoMultipliers
\end{keyword}

\end{frontmatter}


\section{Introduction}

The ``shashlik'' readout technology~\cite{Fessler:1984wa} for sampling calorimeters has been successfully employed in particle physics since more than twenty years~\cite{atoyan1992,Badier:1993zj,Alvsvaag:1998bd,Aphecetche:2003zr,Zoccoli:2000dn,Goebel:2000uf,Dzhelyadin:2007zz}. Shashlik calorimeters are sampling calorimeters in which the scintillation light is readout via wavelength shifting (WLS) fibers running perpendicularly to the converter/absorber plates. This technique offers flexibility in the calorimeter design combined with ease of assembly, good hermeticity and fast time response. In most applications, it represents a cost effective solution compared to crystals or cryogenic liquid calorimeters. The main drawback of this technique is the limitation imposed to the longitudinal segmentation by the optical fiber readout. For each segment, fibers must be bundled and routed toward the photosensors, introducing non-uniform response or the presence of dead zones. As a consequence, shashlik calorimeters allow for high flexibility in the choice of parameters except for the granularity of the longitudinal sampling~\cite{Benvenuti:1999qy}.

The current generation of shashlik calorimeters~\cite{Atoian:2007up,Fantoni:2011zza,Anfimov:2013kka} is readout by compact solid-state photosensors. In particular, the INFN FACTOR Collaboration has demonstrated that Silicon PhotoMultipliers (SiPMs) can be used very effectively for the light readout of these detectors~\cite{willie,jack_maroc}. The compactness of SiPMs and the possibility to use these devices embedded in the bulk of the calorimeters~\cite{collaboration:2010hb} open new possibilities to implement longitudinal segmentation in shashlik devices.

\begin{figure}[!htb]
\centering
\includegraphics[width=0.9\textwidth]{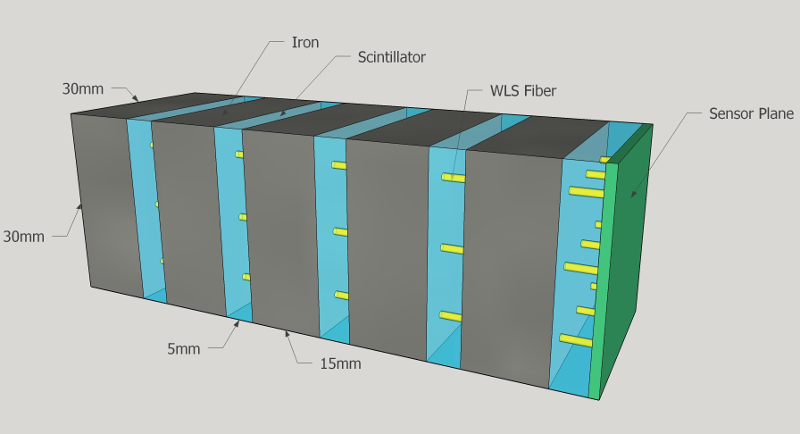}
\caption[]{\label{fig:scentt}Schematics of the ultra-compact module that is being developed by SCENTT. The module corresponds to a 4~$X_0$ sampling and a transverse granularity of 3$\times$3~cm$^2$ (9 fibers per module).}
\end{figure}

In particular, the INFN SCENTT Collaboration is developing an ultra-compact module (Fig.~\ref{fig:scentt}) where every single fiber segment is directly connected to a SiPM and the SiPMs array is hosted on a PCB (Printed Circuit Board) holder that integrates both the passive components and the signal routing toward the front-end electronics (fast digitizers). This setup offers maximum flexibility in the choice of the longitudinal sampling (length of the fiber crossing the scintillator/absorber tiles) and transverse granularity (tile size and number of summed SiPM channels). Applications range from high granularity devices for collider physics employing particle flow algorithms for hadrons~\cite{collaboration:2010hb}, up to beam dump~\cite{Anelli:2015pba} and non-conventional neutrino facilities~\cite{Longhin:2014yta,Berra:2015rgp,enubet}, where the large sizes of the detectors pose stringent requirements on the cost.

In view of the SCENTT R\&D, one of the FACTOR calorimeters was modified to allow for direct fiber-SiPM coupling without bundling and tested with negative charged particles at the CERN PS-T9 beamline. The goals of the test were to demonstrate that direct fiber coupling to SiPMs and summing up of the SiPMs output signal without pre-amplification can replace fiber bundling to a single large area photosensor. In particular, the tests were aimed at checking that direct coupling does not compromise the energy response and linearity achieved in FACTOR. During the test, $e/\pi$ separation, nuclear counter effects (NCE) and the performance of the readout electronics were also investigated.

Sec.~\ref{sec:prototype} and Sec.~\ref{sec:setup} of this paper are devoted to the description of the shashlik prototype and the experimental setup. Sec.~\ref{sec:geant4} presents the GEANT4 simulation of the calorimeter under test. Linearity and energy resolution for electrons are discussed in Sec.~\ref{subsec:lin_ene}, while studies with partially contained hadrons for $e/\pi$ separation, NCE and readout electronics are reported in Secs.~\ref{subsec:epi}, \ref{subsec:NCE} and~\ref{subsec:digitizer}, respectively.

\section{The test calorimeter}\label{sec:prototype}

\begin{figure}[!htb]
\centering \subfloat[\label{fig:calo_1}]{
  \includegraphics[height=0.26\textheight]{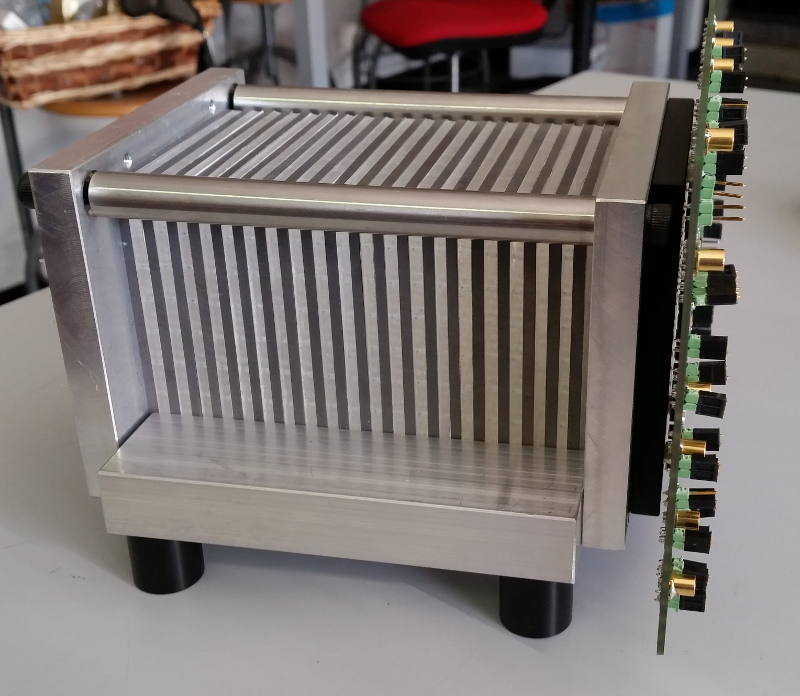}}
\subfloat[\label{fig:calo_2}]{
  \includegraphics[height=0.26\textheight]{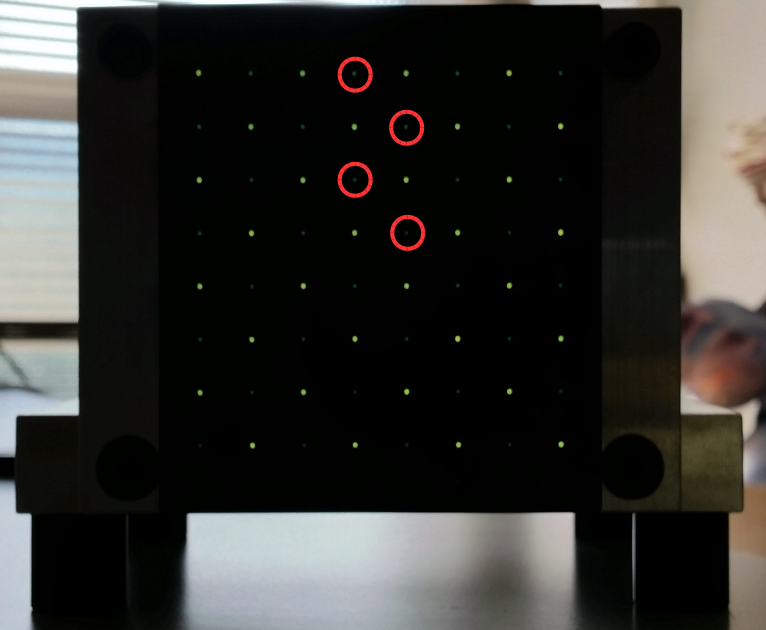}}
\caption[]{\label{fig:calo}Picture of the first module of the calorimeter prototype~\subref{fig:calo_1}; close-up on the WLS fibers~\subref{fig:calo_2} and the long/short alternated pattern: short fibers (some of them highlighted with red circles) are less bright than long ones due to the different ambient light collection efficiency.}
\end{figure}

The shashlik calorimeter prototype employed in this study is a modified version of the calorimeter tested by FACTOR in 2009~\cite{willie}. It is composed of two modules: each module consists of 8$\times$8~cm$^2$ tiles of lead interleaved with plastic scintillator. The thickness of both the lead and scintillator tiles is 3.3~mm and each module groups 20 (lead) + 20 (scint.) tiles. The depth of the module corresponds to $\sim12~X_0$ (Fig.~\ref{fig:calo_1}). Due to the limited amount of SiPMs available for the test only one module per run could be readout: the longitudinal containment of a 1(5)~GeV electromagnetic (EM) shower in a module is 88\%(83\%) (see Sec.~\ref{sec:geant4}). Given the beam energy range (1-5~GeV) and the number of available radiation lengths, a single module with different fiber lengths (see below) was used to evaluate the $e$/$\pi$ discrimination capability of the calorimeter in the region of interest for neutrino physics applications (two longitudinal samplings at $\sim$5 and $\sim10~X_0$ for $E<5$~GeV~\cite{Longhin:2014yta}).

The light produced in the scintillator is collected by 64 0.8~mm diameter Kuraray-Y11 WLS fibers. In the first module, the fibers are divided into two groups with different lengths: the former crosses the whole calorimeter ($12~X_0$) and the latter only half of it ($6~X_0$). The pattern of long and short fibers is shown in Fig.~\ref{fig:calo_2}: the module is readout by one short fiber every 2~cm in both transverse directions interleaved with a long fiber. In this configuration the shower is sampled uniformly by both long and short fibers. Long fibers integrate the scintillator light over $12~X_0$ while short fibers collect the light produced between 6 and $12~X_0$ since the beginning of the shower. For comparison, the second module was readout only with long fibers. This configuration is different from the one that will be implemented in SCENTT (one standalone module every 4~X$_0$) but allows for a test of $e$/$\pi$ separation using just one SiPM holder plane located at the back of the module.

\begin{figure}[!htb]
\centering
\subfloat[\label{fig:pcb}]{
\includegraphics[width=0.65\textwidth]{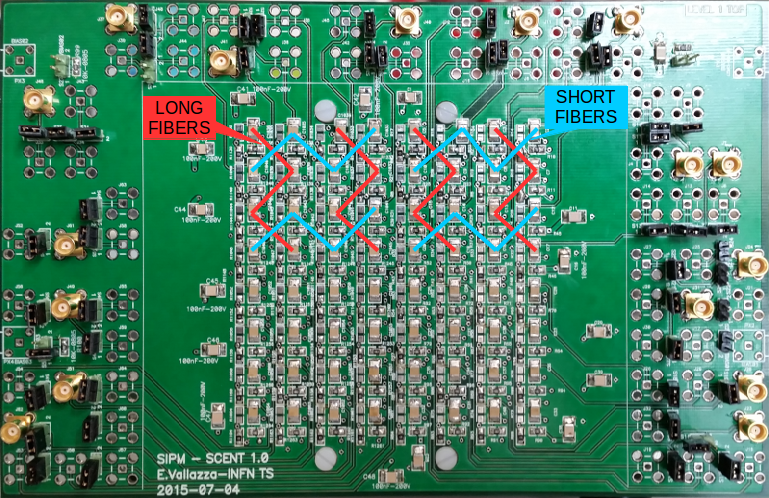}}
\subfloat[\label{fig:sipm}]{
\includegraphics[width=0.25\textwidth]{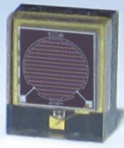}}
\caption[]{\label{fig:electronics}Back-plane of the custom PCB~\subref{fig:pcb} where the SiPMs and the electronic components are hosted (the SiPMs are soldered on the opposite side); close-up of the Advansid 1.13~mm$^2$ circular SiPMs~\subref{fig:sipm} employed in the calorimeter. The red and blue lines in~\subref{fig:pcb} show the group of SiPMs that are summed.}
\end{figure}

The WLS fibers are collected in a plastic holder, which also acts as a mechanical support for the electronic board hosting the SiPMs. 64 SiPMs manufactured by Advansid~\cite{advansid} in a plastic SMD package (Fig.~\ref{fig:sipm}) are soldered on a custom PCB (glass-reinforced epoxy laminate sheet FR4, 4 layers, 1.6~mm total thickness). The mechanical matching between the plastic mask and the PCB ensures that each SiPM is in direct contact with the fiber and is centered along the fiber axis. The technical specifications of the SiPMs employed in this test can be found in Tab.~\ref{tab:sipm}. The electronic board provides:
\begin{itemize}
\item the signal routing through MCX connectors: the 64 SiPMs can be readout independently, or summed together in groups of four, reducing the number of readout channels to 16. The summed configuration can be selected switching a set of jumpers located on the board (see Fig.~\ref{fig:pcb}); the SiPMs are AC coupled to the output with 10~nF capacitors;
\item four independent bias voltages corresponding to four different regions of the board: sensors with the same breakdown voltage were grouped in the same bias zone in order to provide the same overvoltage to all the SiPMs (the SiPMs have been preliminary characterized by the manufacturing company in terms of I-V curves). The precision of the overvoltage equalization is within 0.1~V.
\end{itemize}

The red (blue) signal routing on the PCB (see Fig.~\ref{fig:pcb}) follows an accordion pattern in order to collect the signals from long (short) fibers only. During the beamtest, the SiPM signals were summed in groups of four, reducing the number of total channels to 16.

\begin{table}[!htb]
\centering
\begin{tabular}{|c|c|c|c|c|c|c|}
\hline 
Model       & V$_\mathrm{{BD}}$ & \# of  & Cell area     & Active         & Fill      & PDE \\ 
            & (V)               & cells  & ($\mu$m$^2$)  & Area (mm$^2)$  & factor    &        \\ 
\hline 
ASD-RGB1C-P & $\sim$28          & 673    & 40$\times$40  & 1.13           &$\sim$60\% & 32.5\% \\ 
\hline
\end{tabular}
\caption{\label{tab:sipm}SiPMs features: V$_\mathrm{{BD}}$ refers to the breakdown voltage of the SiPMs, while the Photo Detection Efficiency (PDE) value is obtained at 550~nm.}
\end{table}

\section{Experimental setup}\label{sec:setup}

The shashlik calorimeter prototype was tested at CERN on the PS-T9 beamline using a mixed beam of muons, pions and electrons up to 5~GeV. On the T9 beamline the particles are obtained from the interaction of the primary 24~GeV/c protons of the PS accelerator with a fixed target. The momentum of the extracted beam ranges between 0.5 and 15~GeV/c. The momentum bite $\delta p / p$ can be set in steps of 1\% changing the aperture of the horizontal collimators~\cite{durieu}. The experimental setup on the T9 beamline was composed of:
\begin{itemize}
\item one CO$_2$ Cherenkov detector for the beam particle ID;
\item two silicon strip detectors (SSD)~\cite{prest} for the track reconstruction with a spatial resolution of $\sim$30~$\mu$m;
\item a 10$\times$10~cm$^2$ plastic scintillator for the trigger;
\item the calorimeter modules, hosted inside a darkened metal box.
\end{itemize}

\begin{figure}[!htb]
\centering \subfloat[\label{fig:setup1}]{
  \includegraphics[height=0.26\textheight]{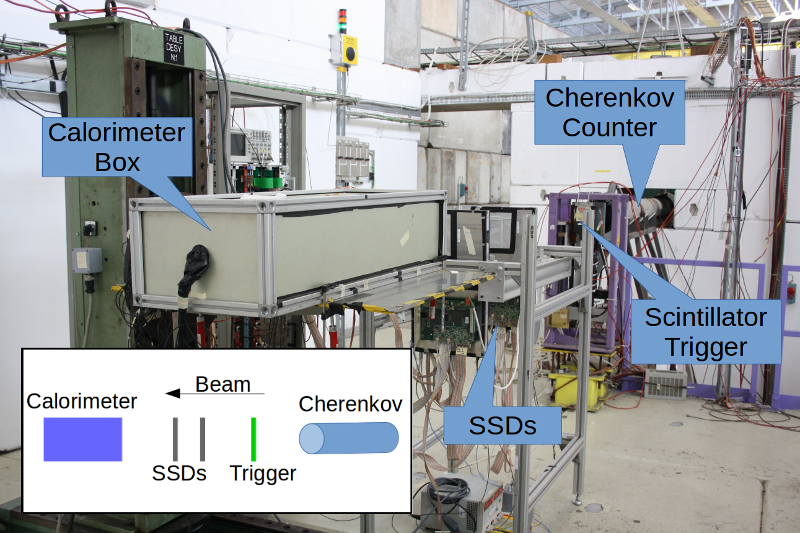}}
\subfloat[\label{fig:setup2}]{
  \includegraphics[height=0.26\textheight]{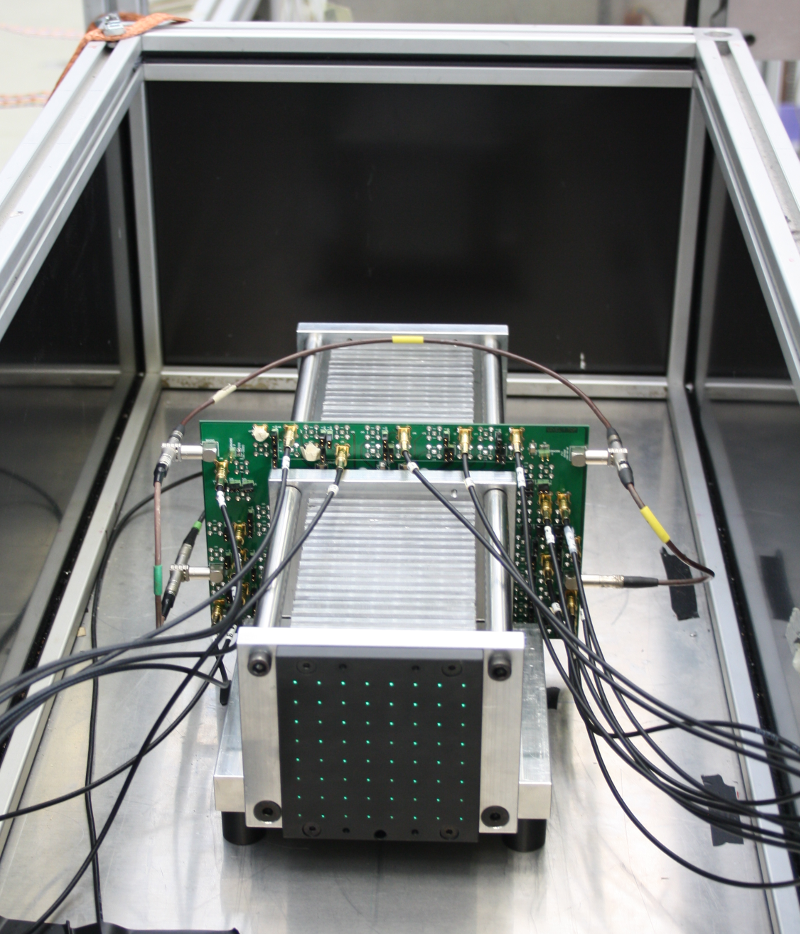}}
\caption[]{\label{fig:setup}The experimental area along the T9 beamline with a sketch of its layout~\subref{fig:setup1}; the calorimeter box with the two modules~\subref{fig:setup2}.}
\end{figure}

The DAQ system is based on a standard VME system controlled by a SBS Bit3 model 620 bridge\footnote{SBS Technologies Inc., US, \texttt{http://www.ge-ip.com}}, optically linked to a Linux PC-system. The calorimeter readout was performed with both a charge integrating ADC (mod. V792 QDC, CAEN) and two 8 channel 500~MS/s, 14-bit waveform digitizers (mod. DT5730 and V1730, CAEN). For the QDC readout, the calorimeter signals were delayed by 120~ns. This readout configuration was used to evaluate the general performance of the calorimeter in terms of linearity and energy resolution (Sec.~\ref{subsec:lin_ene}),  $e$/$\pi$ discrimination capability (Sec.~\ref{subsec:epi}) and possible contributions due to the nuclear counter effects (Sec.~\ref{subsec:NCE}).

In addition, special runs were performed with a waveform digitizer DAQ (Sec.~\ref{subsec:digitizer}). In these runs, a 1024~ns (corresponding to 512 samples) waveform was acquired for each calorimeter channel without any delay line (the digitizer sampling started 60~ns before the trigger signal). In both DAQ modes, the trigger conditions and the silicon strip detector readout were controlled by custom VME boards developed by INFN-Trieste~\cite{jack_maroc}.

\section{GEANT4 simulation}\label{sec:geant4}

\begin{figure}[!htb]
\centering
\includegraphics[width=0.7\textwidth]{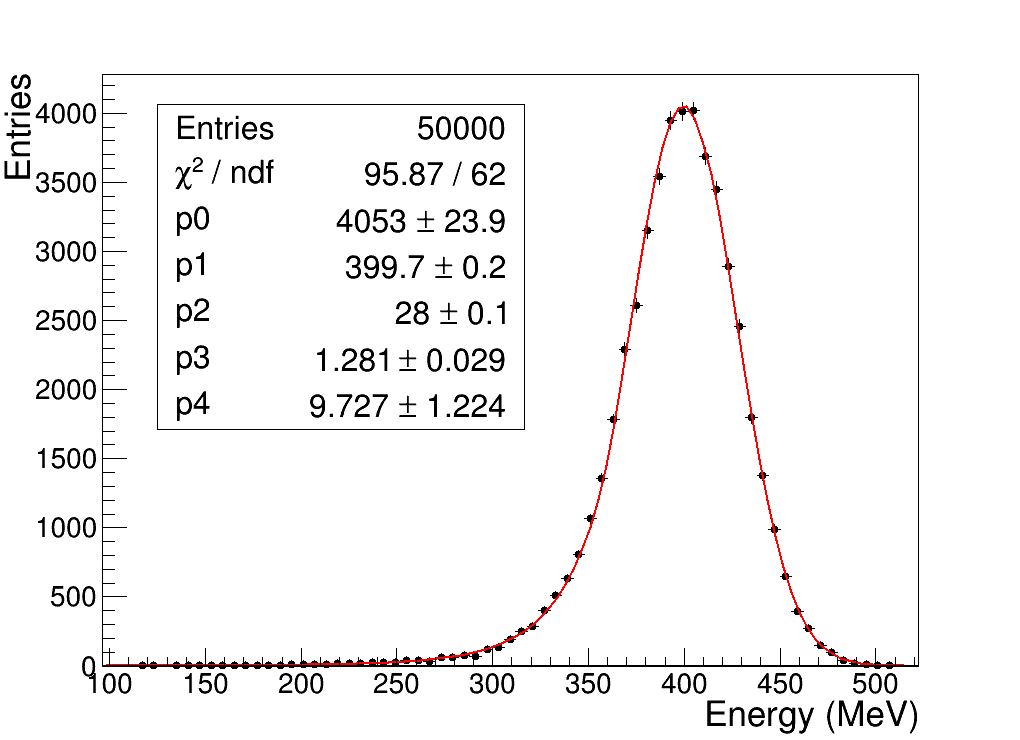}
\caption[]{\label{fig:geant4_cb}(GEANT4 simulation) Energy deposit in the scintillator with an electron beam of 5~GeV with the Crystal Ball fit superimposed.}
\end{figure}

The calorimeter prototype was simulated using the GEANT4~\cite{geant4} simulation package in order to characterize the detector in terms of linearity, energy resolution, energy leakage and $e$/$\pi$ discrimination. The simulation is based on the energy deposited in the scintillator tiles and, hence, it sets the reference for an ideal light readout system. In particular, it does not account for light production and transport in the WLS fibers and for the SiPM photon detection efficiency. The primary particle beam consists of $5\times 10^4$ monochromatic electrons and pions: five different runs have been generated, with an energy of the primary particles in the 1-5~GeV range and momentum parallel to the axis of the calorimeter. Considering its negligible contribution (1.4\%), the momentum dispersion of the T9 beamline is not included in the primary particle generation. The distribution of the particles in the transverse plane is a two dimensional Gaussian ($\sigma$=4~cm) centered on the axis of the calorimeter. Only particles impinging in the $4\times4$~cm$^2$ central area are considered for the analysis. These parameters reproduce the beam size of the CERN PS-T9 beamline as monitored by the SSD during the test and the fiducial area cut employed using the SSD hits. The FTFP\_BERT reference physics list~\cite{FTFPBERT} was used to describe the physical interactions of the electrons and hadrons in the calorimeter. The calorimeter linearity, energy resolution and shower containment were obtained fitting the histograms of the energy deposit inside the scintillator tiles with a Crystal Ball function~\cite{crystalball}(Fig.~\ref{fig:geant4_cb}).

\begin{figure}[!htb]
\centering \subfloat[\label{fig:geant4_res_line}]{
  \includegraphics[width=0.49\textwidth]{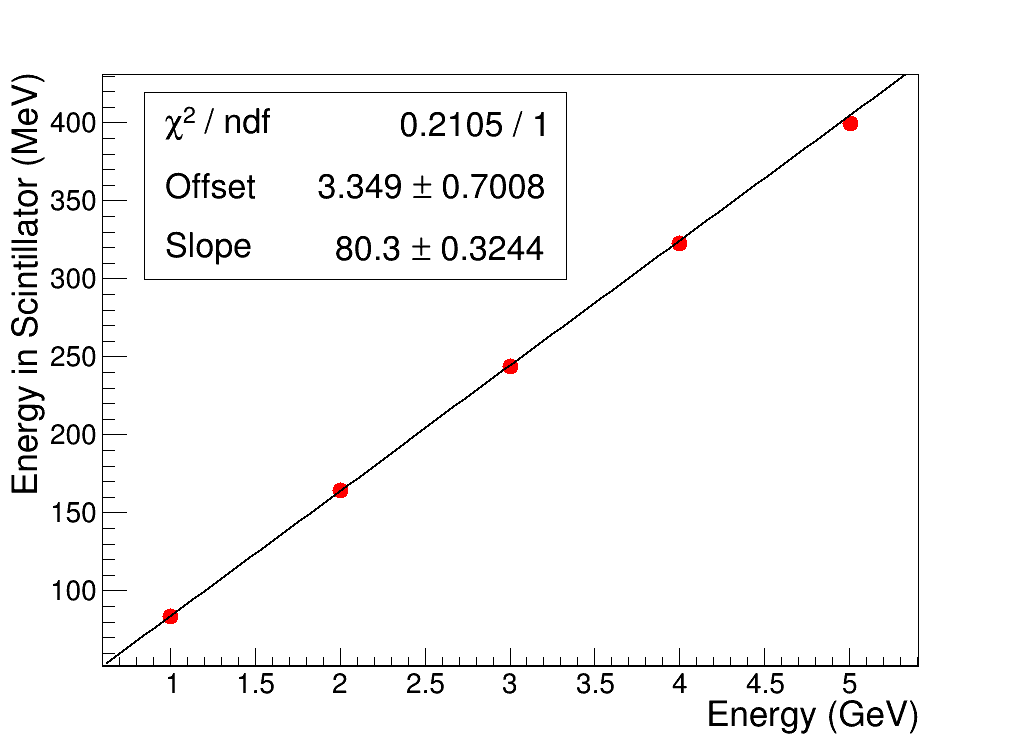}}
\subfloat[\label{fig:geant4_res_res}]{
  \includegraphics[width=0.49\textwidth]{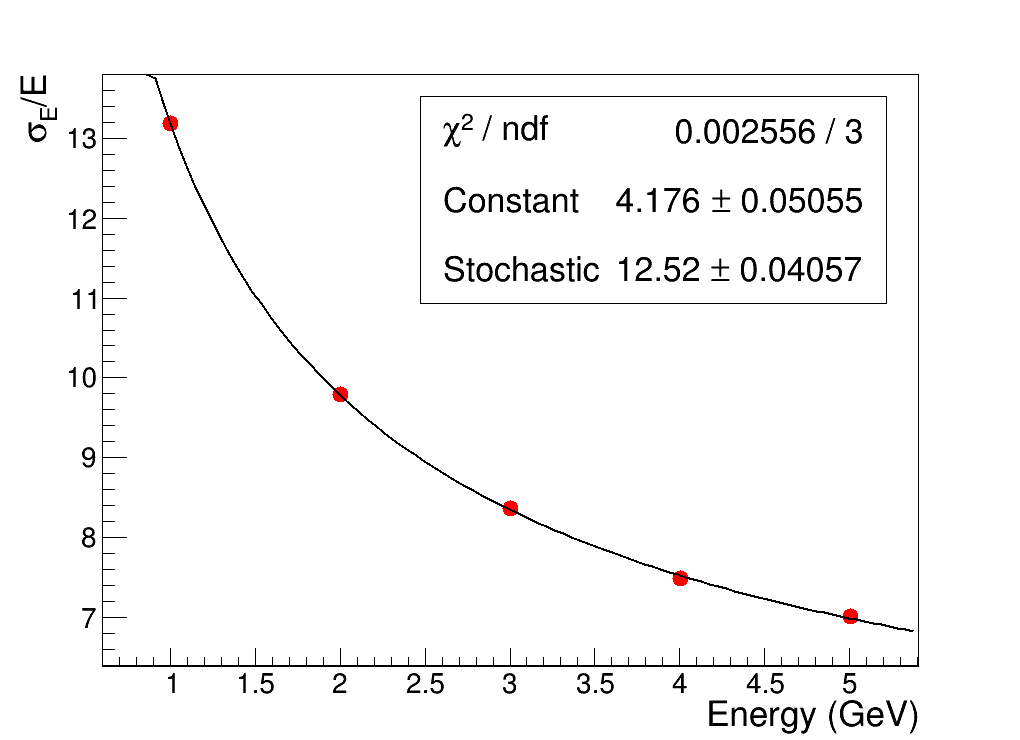}}
\caption[]{\label{fig:calo_res}(GEANT4 simulation) Energy linearity~\subref{fig:geant4_res_line} and energy resolution~\subref{fig:geant4_res_res}. The statistical errors are present, but too small to extrude outside the marker points.} 
\end{figure}

The mean and the $\sigma$ values of the Crystal Ball function were used to test the linearity of the detector response (Fig.~\ref{fig:geant4_res_line}), the shower containment fraction and the energy resolution (Fig.~\ref{fig:geant4_res_res}) as a function of the particle momentum. The energy resolution was fitted with a $\frac{\sigma_E}{E}=\frac{S}{\sqrt{E}} \oplus C$ function, where $S$ and $C$ represent the stochastic and constant terms~\cite{fabjan}, respectively.

As expected, small deviations from the linearity (0.6\% and 1.2\%) are visible above 3~GeV: this is due to the limited size of the calorimeter (12~$X_0$) and the resulting longitudinal energy leakage. The shower containment decreases from 88\% at 1~GeV to 83\% at 5~GeV. The stochastic term obtained from the fit is 12.5\% with a constant term of $\sim$4\% (mostly due to the longitudinal leakage). At low energy, the resolution is dominated by the sampling fraction of the calorimeter.

\begin{figure}[!htb]
\centering \subfloat[\label{fig:geant4_spectrum_merge_2gev}]{
  \includegraphics[width=0.49\textwidth]{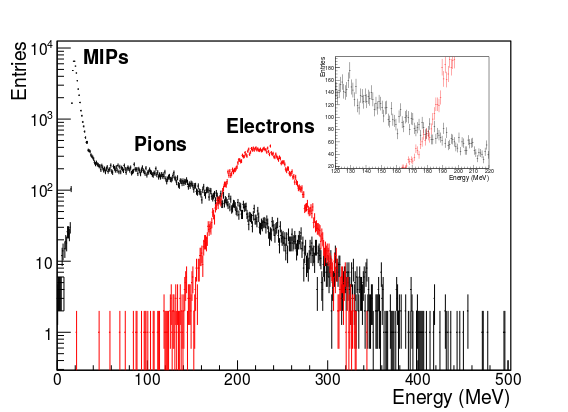}}
\subfloat[\label{fig:geant4_spectrum_merge_5gev}]{
  \includegraphics[width=0.49\textwidth]{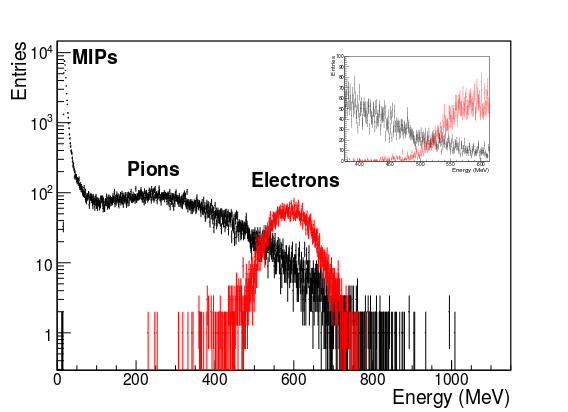}}
\caption[]{\label{fig:geant4_spectrum}(GEANT4 simulation) Energy spectrum for a 2~GeV~\subref{fig:geant4_spectrum_merge_2gev} and 5~GeV~\subref{fig:geant4_spectrum_merge_5gev} $e$/$\pi$ mixed beam. Electron events are shown in red, pions in black; the inserts present a zoom near the $e$/$\pi$ energy deposit transition.}
\end{figure}

In order to estimate the $e/\pi$ separation capability, the energy deposit in all the scintillator tiles (simulating the long fibers light collection) and in the last ten tiles (simulating the short fibers light collection) is considered. In general, a calorimeter can discriminate between electrons and pions exploiting the different pattern of the energy deposit produced by electromagnetic and
hadronic showers. Fig.~\ref{fig:geant4_spectrum} shows the energy deposited in the scintillator tiles by a mixed beam of electrons, pions and kaons (10$^5$ total generated particles) at 2 and 5~GeV. In the simulated beam, the percentage of each particle type corresponds to the nominal PS-T9 beamline composition: 26\% $e^-$, 69\% $\pi^-$ and 5\% $K^-$ at 2~GeV; 5\% $e^-$, 89\% $\pi^-$ and 6\% $K^-$ at 5~GeV. 

The spectrum shows three regions of interest. At low energy, non-interacting pions/kaons deposit a fixed amount of energy (the MIP peak) that can be clearly separated from EM showers. The second peak is located at high energies and it is mostly due to electrons showering inside the calorimeter. The energy deposits between the two peaks are due to hadronic showers of pions and kaons the energy of which partly leaks from the calorimeter. 

\begin{figure}[!htb]
\centering \subfloat[\label{fig:geant4_eff_pur_comp_2gev}]{
  \includegraphics[width=0.49\textwidth]{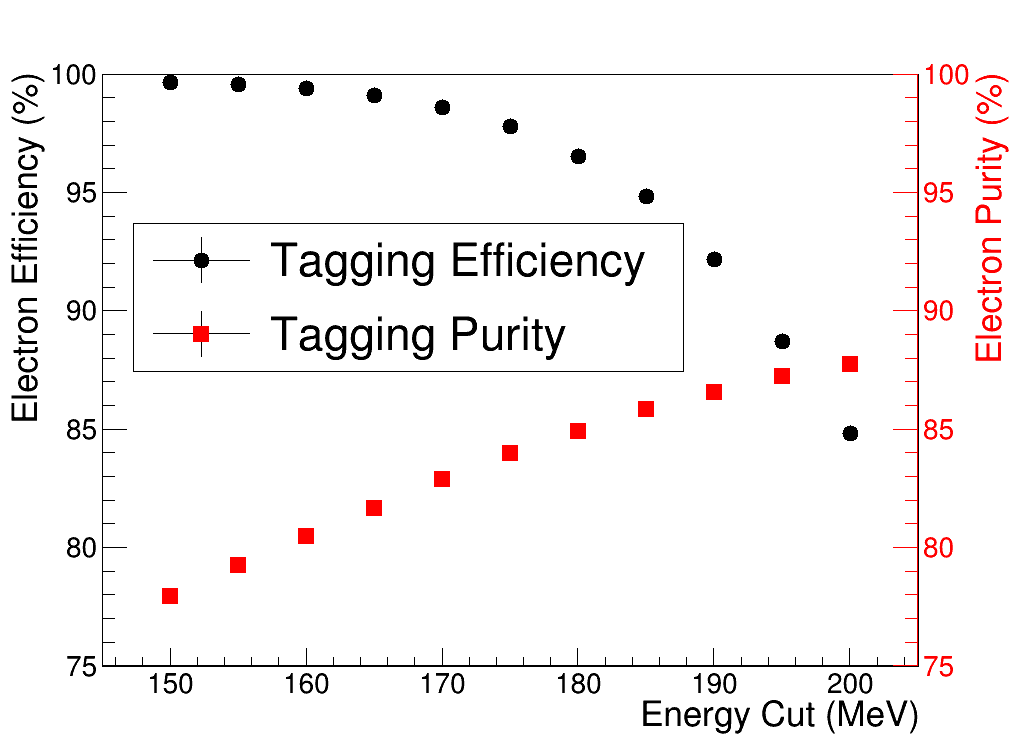}}
\subfloat[\label{fig:geant4_eff_pur_comp_5gev}]{
  \includegraphics[width=0.49\textwidth]{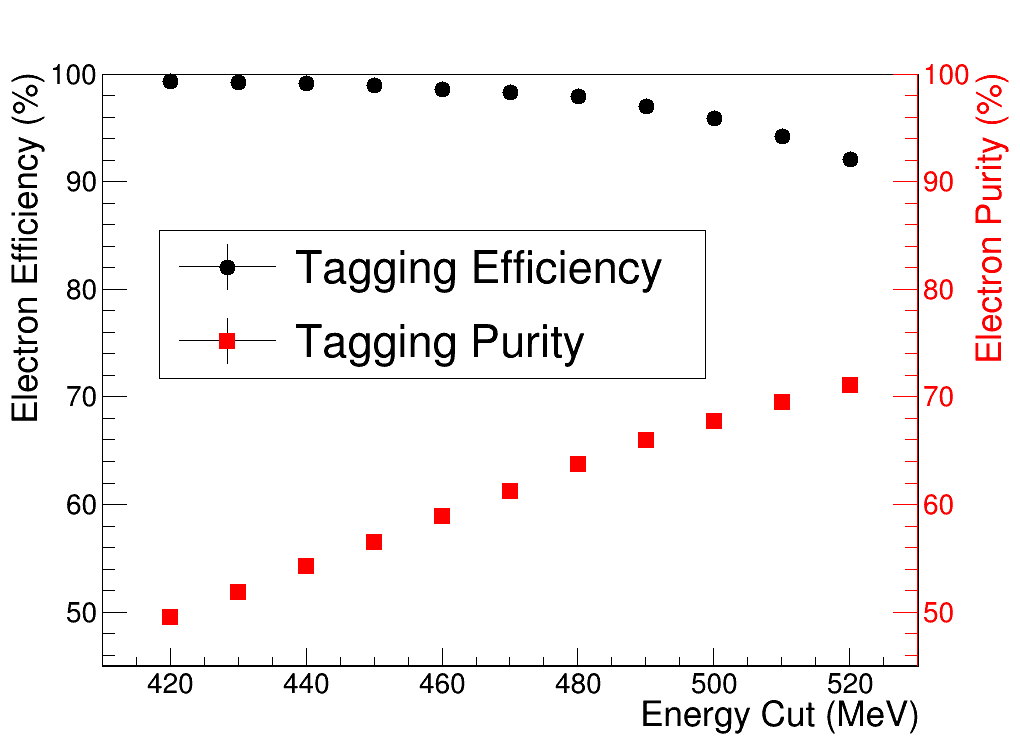}}
\caption[]{\label{fig:geant4_eff_pur_comp}(GEANT4 simulation) Comparison between the efficiency (in black) and the purity (in red) obtained varying the energy cuts with a 2~GeV~\subref{fig:geant4_eff_pur_comp_2gev} and a 5~GeV~\subref{fig:geant4_eff_pur_comp_5gev} $e$/$\pi$ mixed beam.}
\end{figure}

If the energy of the incoming particle is known (either by tracking or by calorimetric measurements of fully contained showers), $e$/$\pi$ separation can be achieved by longitudinal segmentation. Fig.~\ref{fig:geant4_eff_pur_comp} shows a comparison between the efficiency and the purity obtained at 2 and 5~GeV as a function of the energy cut. This cut is applied to the total energy deposited in the scintillator (12~X$_0$ sampling). The purity is defined as the ratio between the number of electrons above the energy threshold and the total number of particles (electrons and pions) with an energy above the threshold. The Monte Carlo evaluation of the purity requires a direct measurement of the beam composition in the proximity of the detector (see Sec.~\ref{sec:results}). The purity estimates reported in the following are based on the nominal composition of the beam and can be used to evaluate the purity gain as a function of the number of longitudinal samples. For a 2~(5)~GeV beam and a threshold corresponding to 98\% electron efficiency, the purity is 84\% (64\%). Additional longitudinal samples can be employed to improve the $e$/$\pi$ separation capability.

\begin{figure}[!htb]
\centering \subfloat[\label{fig:geant4_short_vs_long_2gev}]{
  \includegraphics[width=0.49\textwidth]{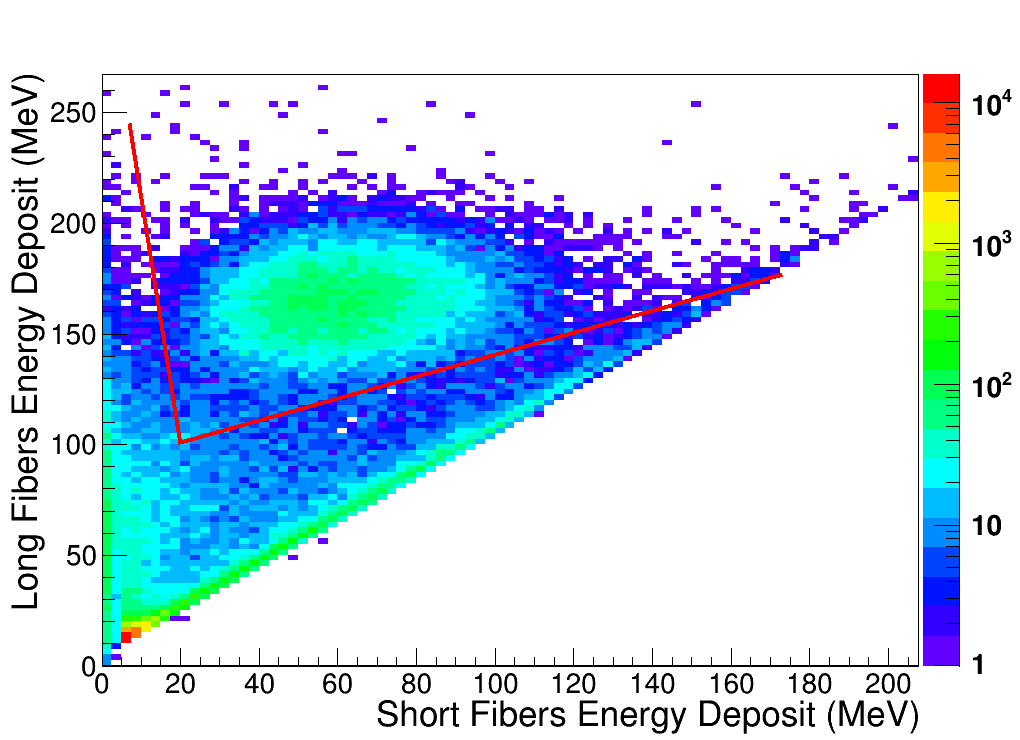}}
\centering \subfloat[\label{fig:geant4_short_vs_long_5gev}]{
  \includegraphics[width=0.49\textwidth]{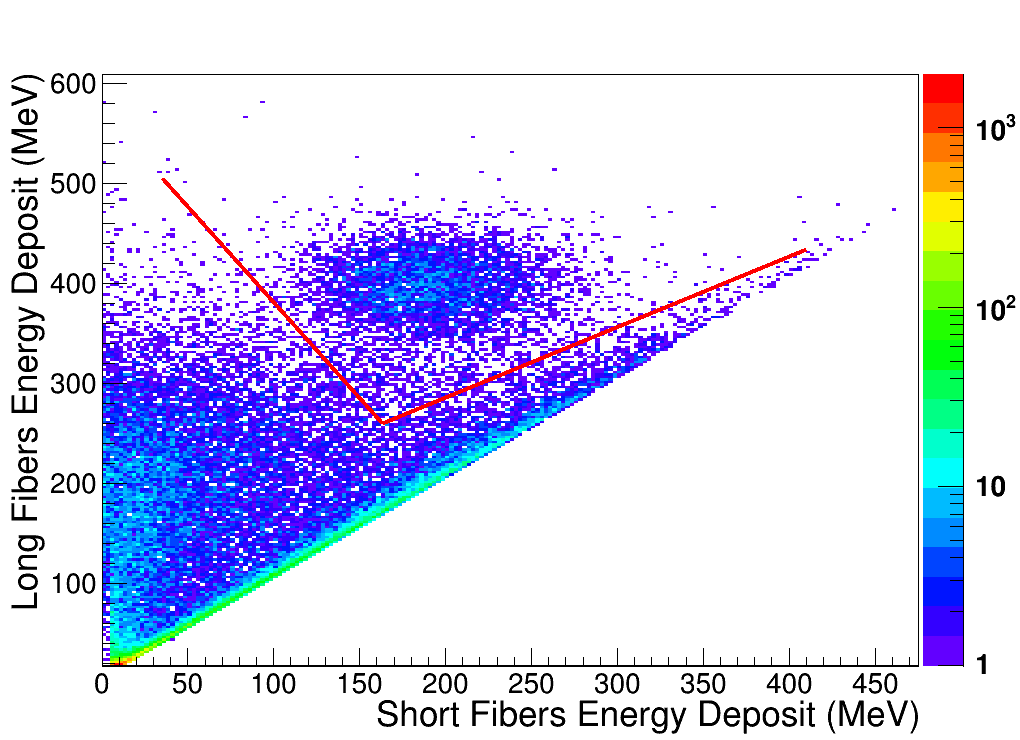}}
\caption[]{\label{fig:geant4_short_vs_long}(GEANT4 simulation) Energy deposited in the scintillator of the whole calorimeter (12~$X_0$ - long fibers) versus energy deposited in the scintillator of the second half of the calorimeter (6~$X_0$ - short fibers) at 2~\subref{fig:geant4_short_vs_long_2gev} and 5~GeV~\subref{fig:geant4_short_vs_long_5gev}. The red lines show the combined energy cut employed for electron identification.}
\end{figure}

Fig.~\ref{fig:geant4_short_vs_long} shows a scatter plot of the energy deposit in all the scintillator tiles ($\mathrm{E_{Long}}$) versus the energy deposit in the last ten tiles ($\mathrm{E_{Short}}$) using a 2~GeV and 5~GeV $e$/$\pi$ mixed beam, mimicking the light readout by the short and long WLS fibers (see Sec.~\ref{sec:prototype}). This is only an approximation of the long (short) fibers signal collection: the actual photon production and collection in long (short) fibers is not simulated.

As expected, a combined selection (red lines in Fig.~\ref{fig:geant4_short_vs_long}) on the energy collected in the long and short fibers improves the $e$/$\pi$ separation: for a 98\% electron efficiency, the purity reaches 92\% (81\%) at 2~(5)~GeV.


\section{Results} \label{sec:results}

The performance of the calorimeter and the $e$/$\pi$ separation capability have been evaluated using the DAQ system with charge integrating readout. The SSD are used to select single track events inside a $4\times4$~cm$^2$ fiducial region centered on the axis of the calorimeter. Electron events are tagged offline combining the SSD and the information from the Cherenkov detector. The distribution of the energy deposited by electrons in the calorimeter is fitted with a Crystal Ball function, whose mean and sigma values are used to estimate linearity and energy resolution.

\subsection{Linearity and energy resolution}\label{subsec:lin_ene}

The linearity and energy resolution of the two shashlik modules is evaluated at different SiPMs overvoltage and tilt angles of the calorimeter with respect to the incident
beam. As shown in Tab.~\ref{tab:conf}, five configurations were considered. The reference setup (n.1 in Tab.~\ref{tab:conf}) is the module instrumented with long fibers only (``long/long'' in Tab.~\ref{tab:conf}) aligned with the direction of the incoming particles. This module was tested with different overvoltages (4 and 5~V) and at a 90~mrad tilt (n.1-3). The second module (``long/short'' corresponding to configurations n.4 and 5) samples only a fraction of the shower at full fiber density and it was used mostly for the $e/\pi$ separation studies (see below).

\begin{table}[!htb]
\centering
\begin{tabular}{|c|c|c|c|}
\hline
Configuration & Module type & SiPM             & Tilt angle \\ 
n.            &             & overvoltage (V)  & (mrad)     \\
\hline
  1           & Long/Long    & 5               & 0           \\ 
\hline
  2           & Long/Long    & 4               & 0            \\ 
\hline
  3           & Long/Long    & 4               & 90           \\ 
\hline  
  4           & Long/Short   & 5               & 0          \\ 
\hline
  5           & Long/Short   & 4               & 90          \\ 
\hline
\end{tabular}
\caption{\label{tab:conf}Calorimeter configurations considered during the data taking.}
\end{table}

\begin{figure}[!htb]
\centering \subfloat[\label{fig:line_norm}]{
  \includegraphics[width=0.49\textwidth]{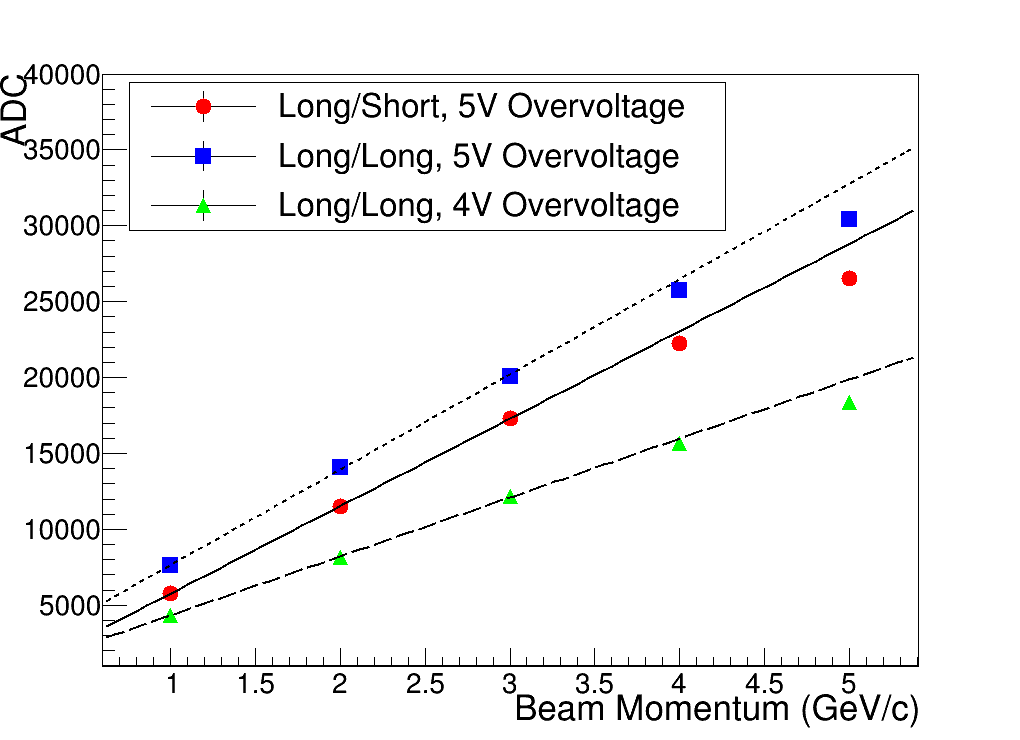}}
\subfloat[\label{fig:line_tilt}]{
  \includegraphics[width=0.49\textwidth]{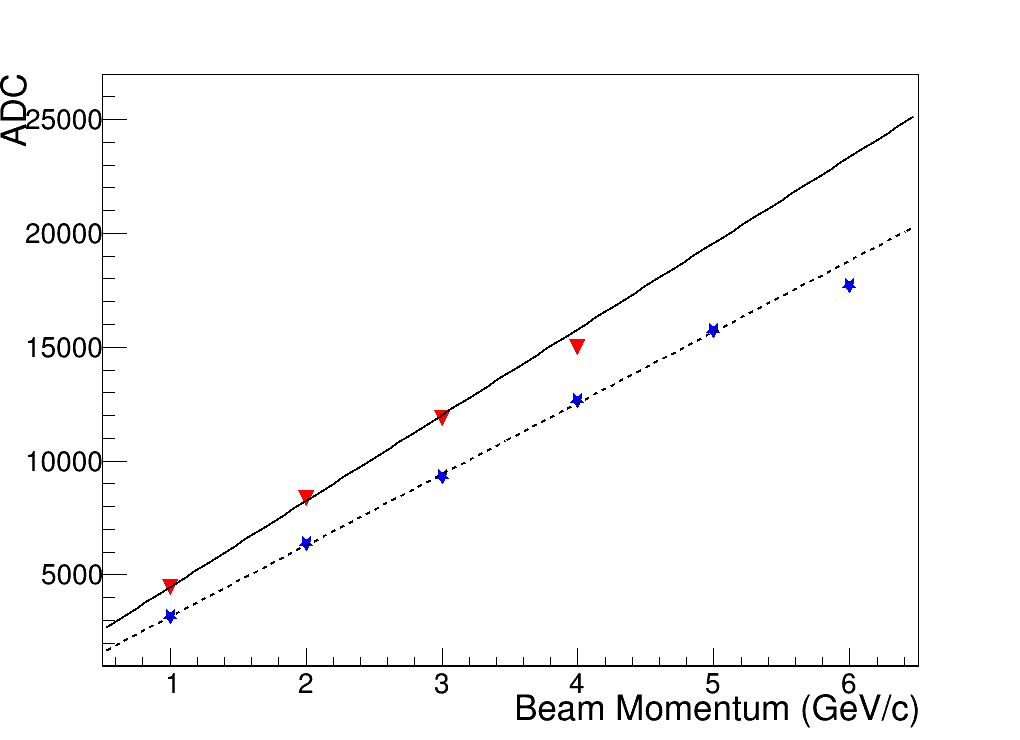}}\\ 
\subfloat[\label{fig:res_norm}]{
  \includegraphics[width=0.49\textwidth]{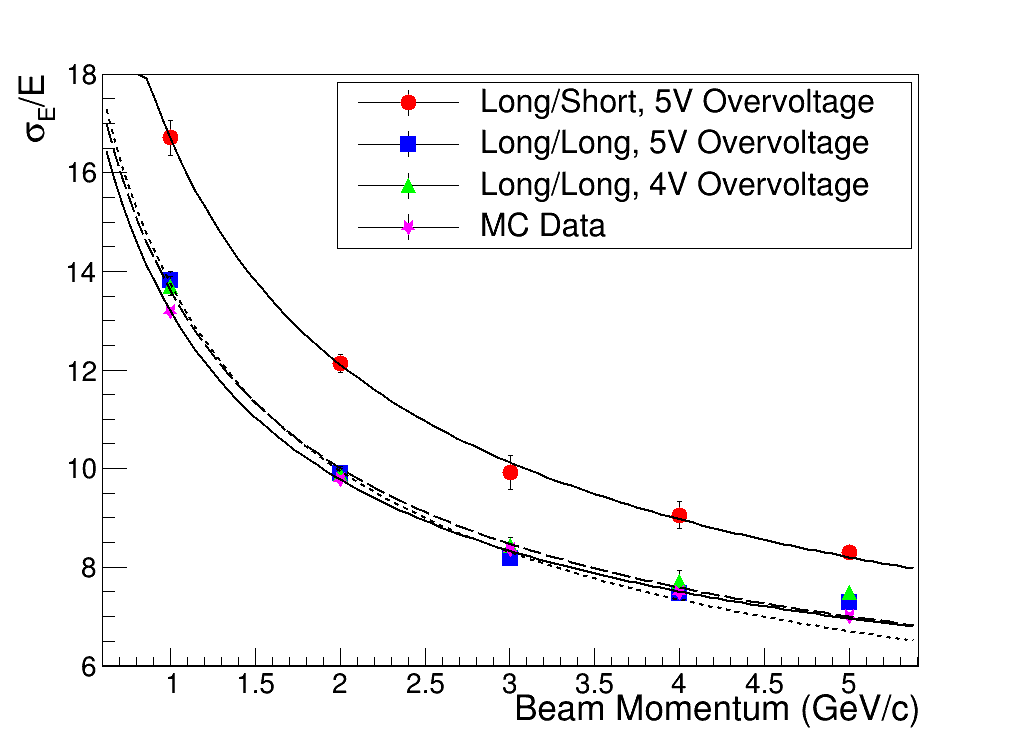}}
\subfloat[\label{fig:res_tilt}]{
  \includegraphics[width=0.49\textwidth]{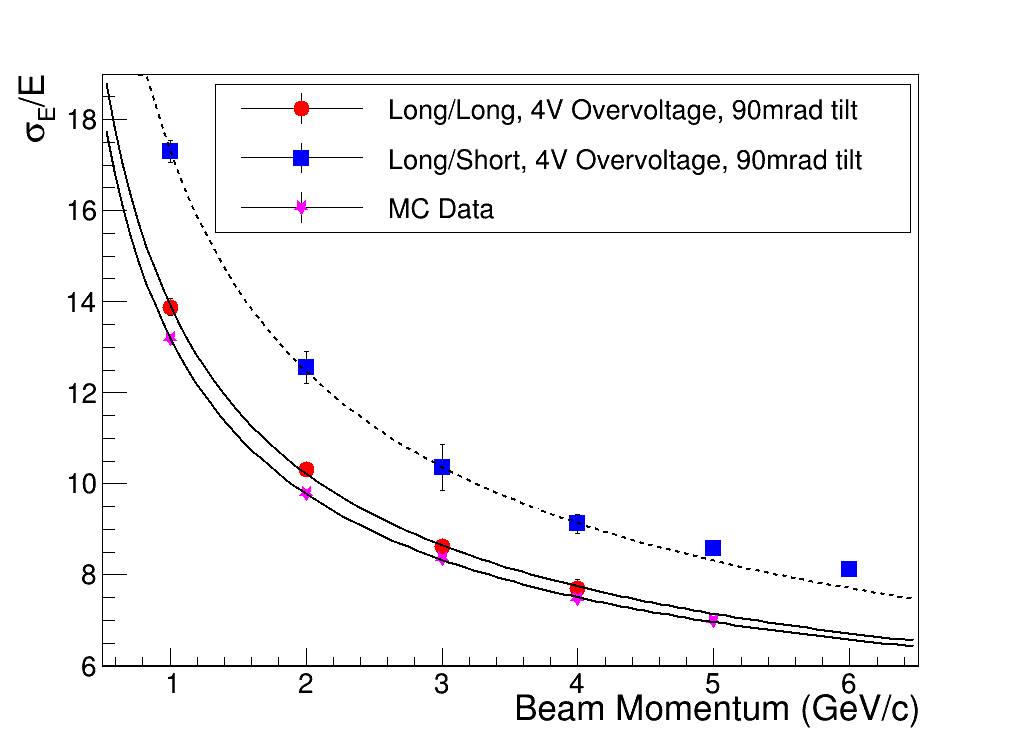}}
\caption[]{\label{fig:calo_general}Energy linearity~\subref{fig:line_norm}\subref{fig:line_tilt} and resolution~\subref{fig:res_norm}\subref{fig:res_tilt} of the calorimeter prototype in different readout configurations. The linearity plots have been fitted in the 1-3~GeV/c momentum range to highlight the deviations above 4~GeV/c; similarly, the 5~GeV/c point has been excluded from the energy resolution fit to decouple the saturation effects due to the SiPM.} 
\end{figure}

The 90~mrad tilt angle was also considered due to its relevance for neutrino physics applications (average positron emission angle from three body kaon decay~\cite{Longhin:2014yta,Berra:2015rgp}). The linearity and resolution results for tagged electrons are presented in Fig.~\ref{fig:calo_general}.

As for the GEANT4 simulation analysis, the energy resolution is fitted with a $\sigma_E/E = S / \sqrt{E} \oplus C$ function. Table~\ref{tab:calo_res_results} summarizes the fitted resolution parameters.

\begin{table}[!htb]
\centering
\begin{tabular}{|c|c|c|}
\hline
Configuration n. & Stochastic     &  Constant        \\
\hline
1                &  $13.4\pm0.3$  &  $3.0\pm0.5$     \\
\hline
2                &  $13.0\pm0.3$  &  $3.9\pm0.5$     \\
\hline
3                &  $13.4\pm0.3$  &  $3.9\pm0.5$     \\
\hline
4                &  $16.3\pm0.5$  &  $3.8\pm0.9$     \\
\hline
5                &  $17.0\pm0.4$  &  $3.4\pm0.9$     \\
\hline
\end{tabular}
\caption{\label{tab:calo_res_results}Stochastic and constant terms as extracted from the energy resolution fits.}
\end{table}

The calorimeter shows saturation effects around 4~GeV/c in all the configurations, which results in deviations from the linearity of the order of 2\% at 4~GeV/c and 8\% at 5~GeV/c. These deviations are larger than the ones obtained in the simulation and they are due to the small number of cells (673) of the SiPMs employed for this test: this saturation effect can be cured with an appropriate choice of the SiPM dynamic range. The energy resolution in the range of interest for neutrino physics applications (few GeVs) is comparable with the resolution predicted by the simulation assuming an ideal light readout system. Moreover, the result is consistent with the energy resolution of shashlik devices with similar lead and scintillator tile thickness operated in the standard fiber bundling scheme~\cite{Alvsvaag:1998bd} and demonstrates the equivalence of the two light readout systems. The performance is similar for particles impinging at small angles (configurations 3 and 5) and is not sensitive to overvoltage variations within 1~V. A deterioration of the energy resolution is visible only for the ``long/short'' module because the calorimeter does not sample uniformly the shower in the proximity of the maximum ($\sim 5 X_0$ at 2~GeV/c).

\subsection{Electron/Pion separation}\label{subsec:epi}

The $e$/$\pi$ discrimination capability has been investigated employing the ``long/short'' module and selecting high purity electron samples with the Cherenkov counter. Due to the strong depletion of the electron component in the T9 beam at high beam momenta ($<$5\% at 5~GeV/c) $e$/$\pi$ separation runs were performed only at 2~GeV/c, i.e. at the peak of the pion background distribution for non conventional neutrino beams~\cite{Longhin:2014yta}.

\begin{figure}[!htb]
\centering
\includegraphics[width=0.7\textwidth]{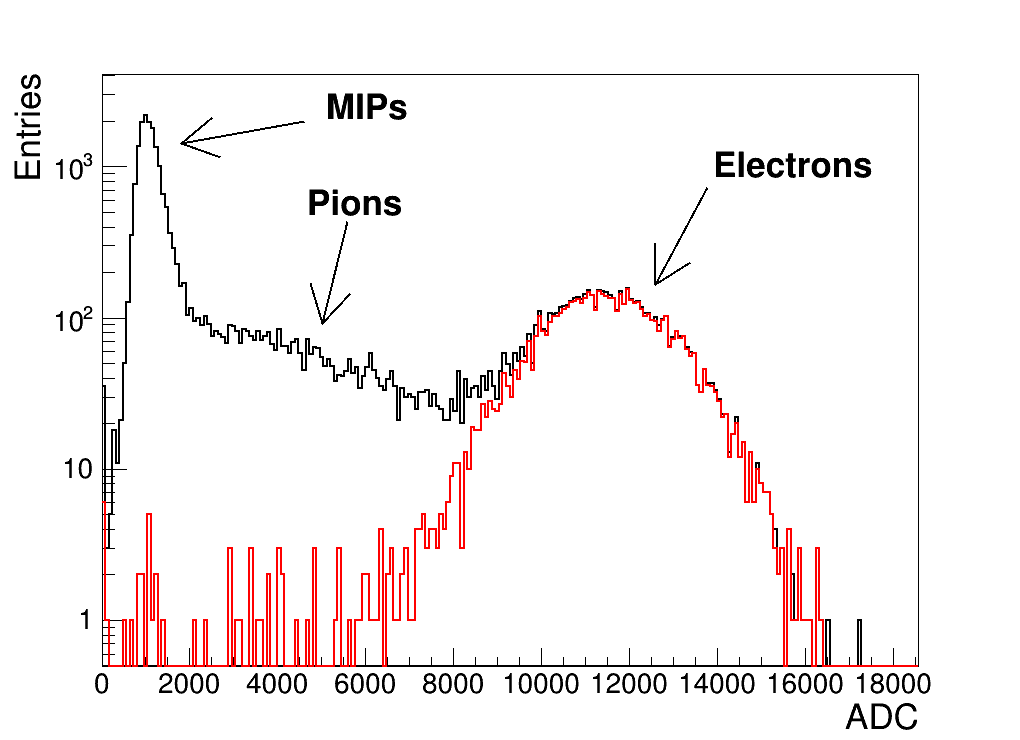}
\caption[]{\label{fig:calo_spectra}Full and Cherenkov-tagged spectrum of the calorimeter exposed to the 2~GeV beam.}
\end{figure}

\begin{figure}[!htb]
\centering
\includegraphics[width=0.7\textwidth]{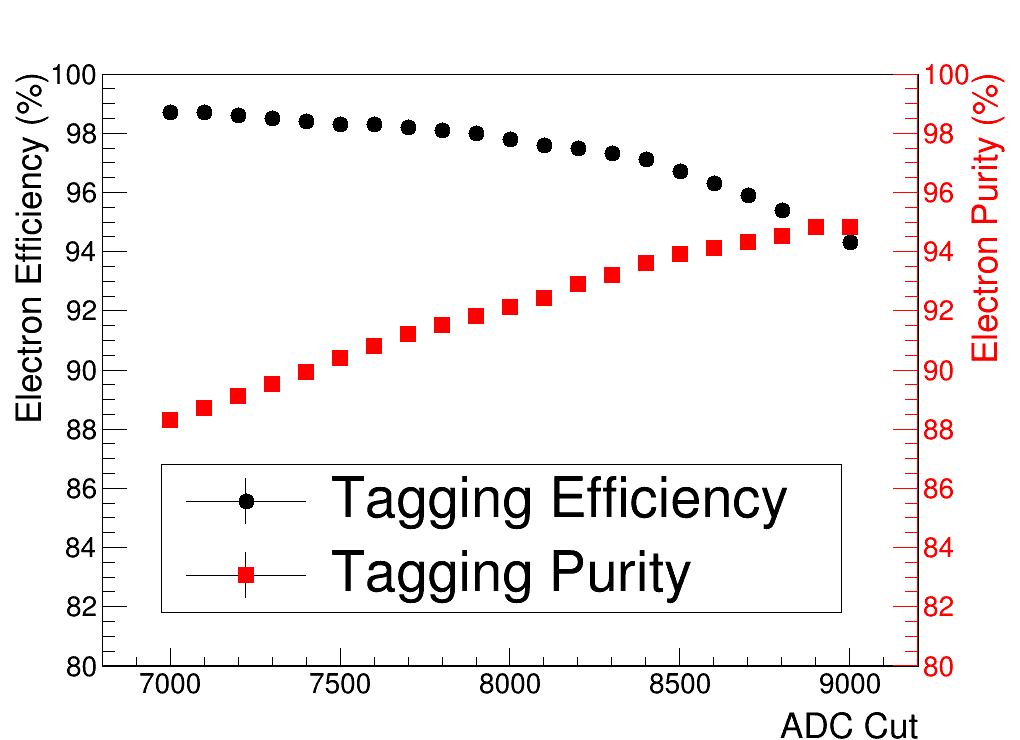}
\caption[]{\label{fig:eff_pur_comp}Comparison between the efficiency (in black) and the purity (in red) obtained varying the energy cut}
\end{figure}

Particle selection is performed using the same criteria as in Sec.~\ref{subsec:lin_ene}. The electron sample (electron efficiency $>$98\%) was selected with the Cherenkov counter, operated with CO$_2$ at a pressure of 1.25~bar. The corresponding energy spectrum (in ADC counts) in the calorimeter is shown in Fig.~\ref{fig:calo_spectra}; the red histogram represents the Cherenkov tagged electron sample, while the full spectrum (black line) includes the MIP peak and interacting hadrons. An energy cut at 7900~ADC counts selects a sample with an electron efficiency of 98\%: the corresponding purity is 91\%. In this case, purity is defined as the ratio between the number of Cherenkov tagged electrons above the energy threshold and the total number of particles (electrons and hadrons) with an energy above the threshold. The efficiency (purity) as a function of the energy threshold is shown in Fig.~\ref{fig:eff_pur_comp}.

\begin{figure}[!htb]
\centering
\includegraphics[width=0.7\textwidth]{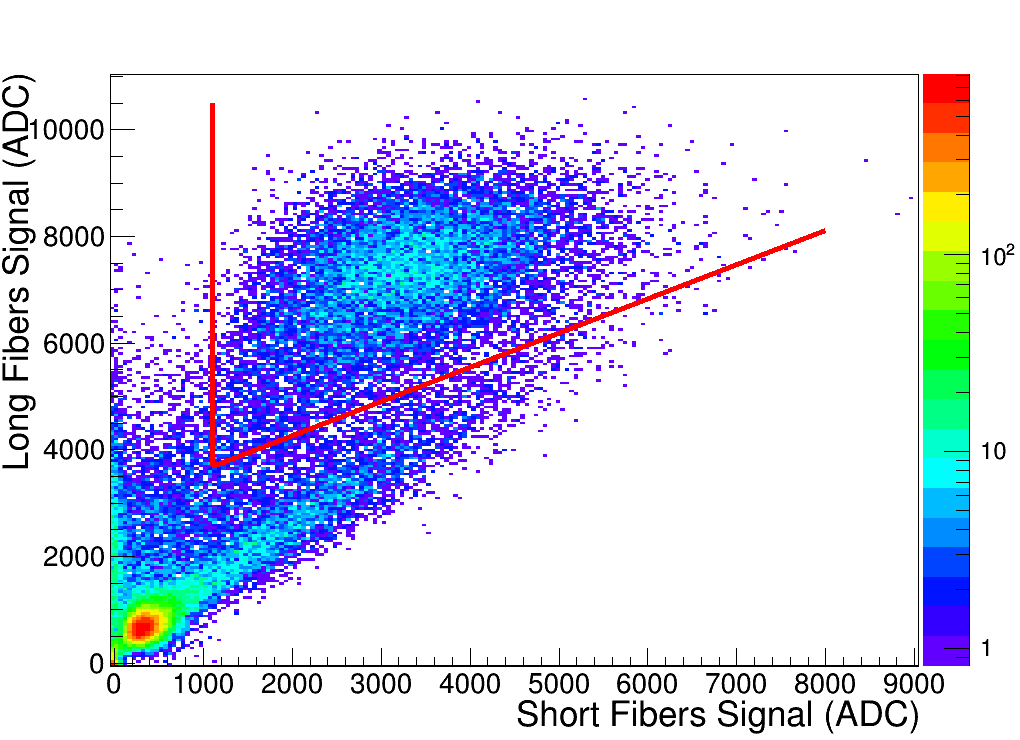}
\caption[]{\label{fig:long_vs_short}
Signal collected by the long versus short fibers at 2~GeV. The red lines show the combined energy cut employed for electron identification.}
\end{figure}

The signal (ADC counts) collected by the long versus short fibers is shown in Fig.~\ref{fig:long_vs_short}. The red lines show the combined cut employed for $e$/$\pi$ separation. For an electron efficiency of 98\%, the combined cut selects a sample with 95\% purity. The additional sampling thus improves the $e$/$\pi$ separation in agreement with the expectation of Sec.~\ref{sec:geant4}. 

A direct comparison of these results with the GEANT4 expectations is limited by the uncertainty on the beam composition. In fact, a full characterization of the beam composition at the proximity of the detector was not possible since only one Cherenkov counter was available. In addition, pressure scans were performed only in a limited range. We hence estimated the systematic uncertainty on the beam composition (mostly due to halo muons and hadron decays) from the ratio between the MIP and the total particle rate in the sample of events with no signal in the Cherenkov counter. The corresponding uncertainty on the purity computed with the long/short information is 3\%.

\subsection{Nuclear counter effects}\label{subsec:NCE}

\begin{figure}[!htb]
\centering
\includegraphics[width=0.7\textwidth]{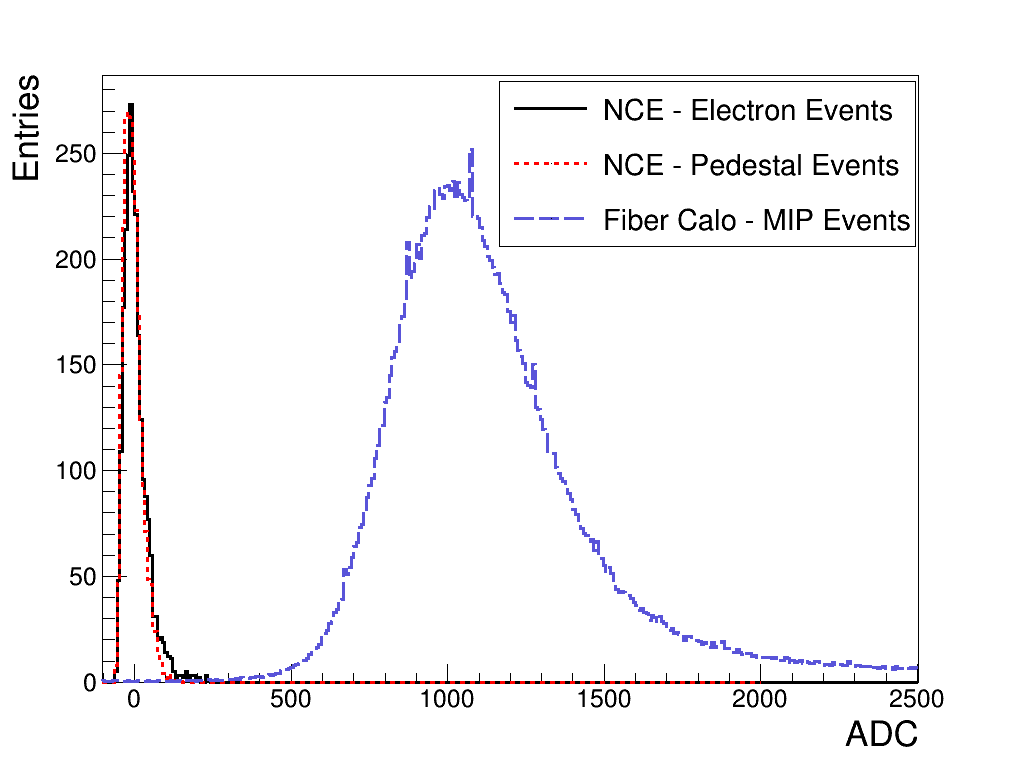}
\caption[]{\label{fig:nuclear_counter_effect}Comparison between the spectra recorded in the special run without WLS fibers (black and red lines - see text for details) and a standard run.}
\end{figure}

Nuclear counter effect (NCE) is the extra amount of charge produced in the photosensor by a particle directly hitting it, on the top of the charge produced by the scintillation light. This direct ionization
effect is of relevance for photodiodes~\cite{Ueno:1997cw}; it has also been observed for APDs in the CMS ECAL calorimeter~\cite{Dittmar:1999uka}, where contributions of direct ionization from particles produced in the hadronic showers~\cite{Zurcher:2000jg,Mao:2011zzc} may affect the $e/\pi$ separation. NCEs are expected to be negligible in SiPMs because these sensors are pixelated and operated in Geiger mode.  When a ionizing particle crosses the SiPM and deposits part of its energy, it triggers a single cell discharge at most, mimicking the signal of only one optical photon~\cite{renker}. No evidence of NCEs were reported by FACTOR where most of the studies were done with electrons and, due to the standard bundling and routing layout, the SiPMs were located far from the core of the shower. Similarly, in the calorimeter studied in this paper the pion samples selected in Sec.~\ref{subsec:epi} do not show distortions that can be traced to fully saturated SiPMs. To test in a direct manner NCE in the present light readout scheme, a dedicated run at 5~GeV/c has been performed. In this special run, the SiPMs were mounted on a calorimeter module with no WLS fibers. Since the field of view of the SiPM towards the scintillator tiles is negligible and the QE of the sensor is mismatched with respect to the emission spectrum of the plastic scintillator, the only expected contribution is from NCE. Fig.~\ref{fig:nuclear_counter_effect} presents the energy spectrum in ADC counts obtained in the NCE run. Two event subsets have been selected:
\begin{itemize}
 \item electron events (black line): electrons tagged with the Cherenkov detector, impinging on the calorimeter. They produce EM showers, thus should maximize NCE signals;
 \item pedestal events (red dashed line): particles that are outside the calorimeter. They should not produce any signal in the calorimeter, and the corresponding energy spectrum can be considered as the electronic chain pedestal.
\end{itemize}

As can be seen, the NCE electron peak is consistent with the pedestal peak and no contributions extend to the MIP region ($\sim$ 1300~ADC counts, blue dashed line) recorded in the same conditions (SiPMs overvoltage, calorimeter position and beam energy) but using a standard module with WLS fibers. This result is consistent with the expectation of negligible NCE in light readout systems based on SiPMs. In the SCENTT setup (Fig.~\ref{fig:scentt}) we expect a factor three larger charged particle rate at 5~GeV since the photosensors are located in the proximity of the maximum of the shower while in this setup they are positioned at 12~$X_0$. Still, from these measurements it is clear that NCE will not represent a relevant source of noise even in the MIP signal region.

\subsection{DAQ based on waveform digitizers} \label{subsec:digitizer}

The application of this light readout system for neutrino physics, as the ones that are being investigated in~\cite{Longhin:2014yta,Berra:2015rgp,enubet}, requires a triggerless acquisition where the signal of the summed SiPMs is continuously sampled during the beam extraction spill. The DAQ system will hence be based on waveform digitizers and sampling will be extended up to 10~ms. In this test, the calorimeter was readout for $\sim$1~$\mu$s (each calorimeter channel is sampled 512 times every 2~ns) with commercial waveform digitizers (CAEN DT5730 and V1730) in order to test data analysis algorithms for the SiPMs waveforms and compare the energy resolution results for fully sampled (500~ms, 14~bit) and downsampled (250~ms, 12~bit) waveforms with the ones obtained with the QDC. The downsampled waveforms are obtained in the offline analysis halving the number of samplings and omitting the last two bits of data.

\begin{figure}[!htb]
\centering
\subfloat[\label{fig:signal_vs_derivative_single}]{
\includegraphics[width=0.7\textwidth]{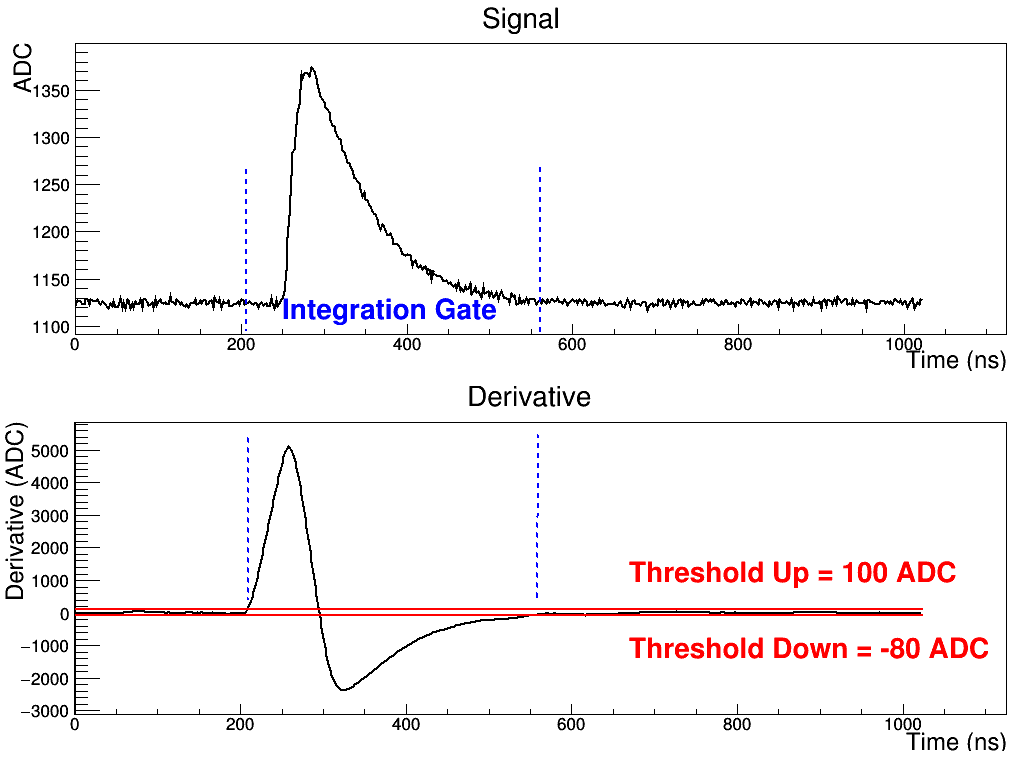}}\\
\subfloat[\label{fig:signal_vs_derivative_multi}]{
\includegraphics[width=0.7\textwidth]{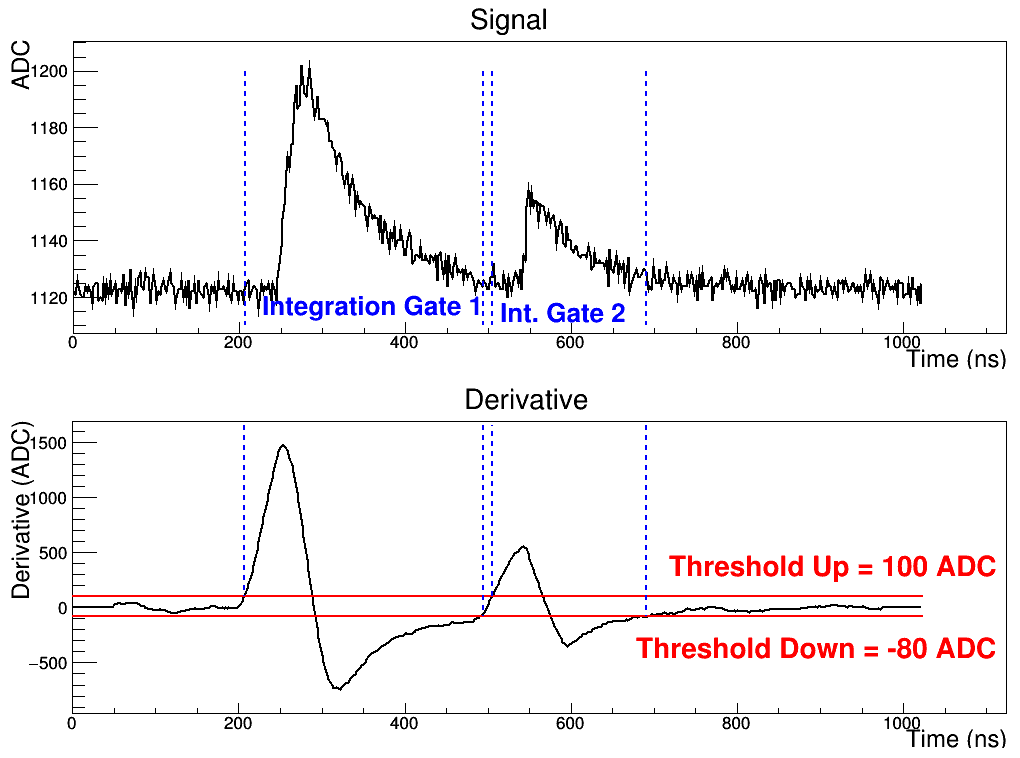}}
\caption[]{\label{fig:signal_vs_derivative}Inverted waveform and corresponding point-by-point smoothed derivative with thresholds and integration gates for a single SiPM signal peak~\subref{fig:signal_vs_derivative_single} and for two adjacent signal peaks~\subref{fig:signal_vs_derivative_multi}. The change of slope of the derivative appears in advance with respect to the actual rising of the signal due to the moving average window algorithm.}
\end{figure}

For each waveform, the smoothed derivative $\delta_i$ of the $i$-th waveform point is computed as:
\begin{equation}
 \delta_i(N)=\sum\limits_{k=1}^{N} s_{i + k} - \sum\limits_{k=1}^{N} s_{i - k}
\end{equation}
where $s_i$ is the waveform value in the $i$-th point and $N$ is the number of points used for the moving average window. A SiPM pulse is tagged when the derivative exceeds for the first time a given positive
threshold (100~ADC counts in Fig.~\ref{fig:signal_vs_derivative_single}). The threshold crossing generates the start of the integration gate; the gate is stopped when the derivative crosses from below a negative threshold (corresponding to -80~ADC in Fig.~\ref{fig:signal_vs_derivative_single}). This algorithm offers high frequency noise filtering (smoothing) and is well suited to resolve multiple peaks in standard SiPM output signals (see Fig.~\ref{fig:signal_vs_derivative_multi}). The sensitivity to the peaks of a given height can be tuned adjusting the upper and lower threshold values. The waveform baseline is computed peak by peak exploiting the waveform integral used for the calculation of the derivative and it is subtracted in the peak integral calculation.

For each event, the 16 waveforms of the calorimeter signals are analyzed. The peaks are identified and integrated in order to compute the total energy deposit. The event selection criteria (fiducial area and
Cherenkov cuts) are the same as for the QDC analysis described above.

In Table~\ref{tab:res_comparison} the energy resolution obtained with the waveform digitizer DAQ (with the original and downsampled waveforms) and the QDC DAQ are compared. The calorimeter used for the comparison is the long/short longitudinally segmented module and the SiPMs are biased with a 4~V overvoltage.

\begin{table}[!htb]
\centering
\begin{tabular}{|c|c|c|c|c|}
\hline
Beam      & 500~MS/s  & 250~MS/s  & QDC        & QDC (4~V OV,  \\
Momentum  & 14-bit    & 12-bit    & (5~V OV)   & 90~mrad Tilt)  \\
\hline
1~GeV/c     & 17.0\%    & 17.2\%    & 16.7\%     & 17.3\%        \\
\hline
2~GeV/c     & 13.4\%    & 13.3\%    & 12.1\%     & 12.6\%        \\
\hline
3~GeV/c     & 10.8\%    & 10.8\%    & 9.9\%      & 10.4\%        \\
\hline
\end{tabular}
\caption{\label{tab:res_comparison}Comparison between the energy resolution values at different energies, using the waveform digitizer based and the QDC based DAQs. The QDC results have been obtained using a slightly larger overvoltage with respect to the waveform digitizer results.}
\end{table}

The energy resolution computed with the waveform digitizer analysis is the same using both the original and the downsampled (250~MS/s, 12-bit) waveforms.  Moreover, the results are compatible with the previous
ones obtained with the QDC based DAQ.

\section{Conclusions}

A calorimeter light readout system based on direct fiber-photosensor coupling without fiber bundling has been tested and characterized on the CERN PS-T9 beamline. This novel system allows for the construction of compact sub-modules that ease significantly the longitudinal segmentation of shashlik calorimeters and will be studied in the framework of the SCENTT R\&D program. In this paper we characterized two shashlik modules where each fiber is directly connected to a SiPM and the SiPMs array is hosted on a PCB holder that integrates both the passive components and the signal routing toward the front-end electronics. One of the two modules was equipped with fibers of different lengths to sample different parts of the shower and improve the $e$/$\pi$ discrimination capability. The energy resolution obtained with this readout scheme ($13\% / \sqrt{E (GeV)} \oplus 3\%$) is comparable to the standard fiber bundling readout and, therefore, the new compact scheme with the photosensors embedded in the bulk of the calorimeter fully retains the performance of the device. Deviations from linearity were observed at beam momenta $>4$~GeV/c and are attributed to the limited dynamic range of the SiPMs used in the test. At 98\% electron efficiency, the sample purity increases from 91\% to 95\% when employing both long and short fibers for electron identification.

The same results can be obtained replacing the QDC DAQ with a 4~ns 12 bit waveform digitizer, as requested for neutrino physics applications. Finally, nuclear counter effects have been tested in a direct manner removing the WLS fibers from one of the modules and keeping the SiPM in the bulk of the module. As expected from the pixelated structure of the SiPM, no visible effects were observed.

\section{Acknowledgments}
This project has received funding from the European Union’s Horizon 2020 Research and Innovation programme under Grant Agreement no. 654168. The authors gratefully acknowledge CERN and the PS staff for successfully operating the East Experimental Area and for continuous supports to users.  We specially thank Henric Wilkens and Lau Gatignon for help and suggestions during the data taking on the PS-T9 beamline.

\end{document}